\documentclass[twocolumn,preprintnumbers,amsmath,amssymb,superscriptaddress]{revtex4}
\usepackage{graphicx}
\usepackage{dcolumn}
\usepackage{bm}
\usepackage{soul}
\usepackage{color}
\usepackage{epstopdf}
\usepackage[version=3]{mhchem}
\usepackage{lipsum}
\usepackage[outercaption]{sidecap}
\usepackage{floatrow}

\begin{document}

\title{Elastically Collective Nonlinear Langevin Equation Theory of Dynamics in Glass-Forming Liquids: Transient Localization, Thermodynamic Mapping and Cooperativity 
}
\author{Anh D. Phan}
\affiliation{Department of Physics, University of Illinois, 1110 West Green St, Urbana, Illinois 61801, USA}
\author{Kenneth S. Schweizer}
\affiliation{Department of Materials Science and Chemistry, Frederick Seitz Materials Research Lab, University of Illinois at Urbana-Champaign}
\email{kschweiz@illinois.edu}

\date{\today}

\begin{abstract}
We analyze multiple new issues concerning activated relaxation in glassy hard sphere fluids and molecular and polymer liquids based on the Elastically Collective Nonlinear Langevin Equation (ECNLE) theory. By invoking a high temperature reference state, a near universality of the apparent dynamic localization length scale is predicted for liquids of widely varying fragility, a result that is relevant to recent simulation studies and quasi-elastic neutron scattering measurements. In contrast, in the same format strongly non-universal behavior is found for the activation barrier that controls long time relaxation. Two measures of cooperativity in ECNLE theory are analyzed. A particle-level total displacement associated with the alpha relaxation event is found to be only of order 1-2 particle diameters and weakly increases with cooling. In contrast, an alternative cooperativity length is defined as the spatial scale required to recover the full barrier and bulk alpha time. This length scale grows strongly with cooling due to the emergence in the deeply supercooled regime of collective long range elastic fluctuations required to allow local hopping. It becomes very large as the laboratory $T_g$ is approached, though is relatively modest at degrees of supercooling accessible with molecular dynamics simulation. The alpha time is found to be exponentially related to this cooperativity length over an enormous number of decades of relaxation time that span the lightly to deeply supercooled regimes. Moreover, the effective barrier height increases almost linearly with the growing cooperativity length scale. An alternative calculation of the collective elastic barrier based on a literal continuum mechanics approach is shown to result in very little change of the theoretical results for bulk properties, but leads to a much smaller and less temperature-sensitive cooperativity length scale.
\end{abstract}

\maketitle


\section{Introduction}
The construction of a quantitative, predictive, force-level theory of activated glassy structural relaxation at the level of atoms or molecules remains a grand challenge in statistical mechanics [1,2,3]. Recently, Mirigian and Schweizer formulated and applied a force-based dynamical theory that relates thermodynamics, structure and activated relaxation for colloidal suspensions [4,5], supercooled molecular liquids [4,6] and polymer melts [7] -- the “Elastically Collective Nonlinear Langevin Equation” (ECNLE) theory. Quantitative tractability for real materials is achieved based on an a priori mapping of chemical complexity [4] to a thermodynamic-state-dependent effective hard sphere fluid using experimental equation-of-state data. The basic relaxation event involves coupled large amplitude cage-scale hopping and a long range but low amplitude collective elastic distortion of the surrounding liquid, resulting in two inter-related, but distinct, barriers. The elastic barrier becomes very important in the deeply supercooled regime and grows much faster with cooling than the local cage barrier. 

The initial formulation of ECNLE theory for rigid molecules is based on a quasi-universal mapping, is devoid of fit parameters, has no divergences at finite temperature or below random close packing, and accurately captures the alpha relaxation time over 14 decades [4,5]. Extension to polymer liquids is based on a disconnected Kuhn segment model [7]. To capture the wide variation of fragility in polymer melts, non-universality was introduced motivated by the system-specific nature of the nm-scale conformational dynamics required for segmental hopping [8]. Good results have been demonstrated for $T_g$, fragility and the temperature dependent segmental relaxation time.

ECNLE theory has also been extended and applied to other problems: spatially heterogeneous relaxation in free standing thin films [9,10,11], segmental relaxation in polymer nanocomposites [12], attractive glass and gel formation in dense sticky colloidal suspensions [13], the effect of random pinning in dense liquids [14], penetrant diffusion in supercooled liquids and glasses [15,16], and activated relaxation in dynamically-asymmetric 2-component mixtures [17].  

In this article, we revisit the basics of ECNLE theory of 1-component liquids to further establish it physical picture and address new questions. After a brief review of key technical aspects in section II, new numerical studies are presented in section III that explore a possible universality of the dynamic transient localization length, and alternative perspectives of the temperature-dependent barrier and effective volume fraction are also studied. Section IV analyzes the magnitude and temperature dependence of the particle-level cooperative displacement of the alpha process and an alternative measure of a growing cooperativity length scale. The latter is shown to be strongly correlated with the alpha time. An alternative continuum mechanics analysis of the elastic barrier and its consequences on the alpha time and cooperativity length scale is presented in Section V. The article concludes in Section VI with a discussion.

\section{ENCLE Theory and Chemical Mapping}
As relevant background, the present state of bulk liquid ECNLE theory is briefly reviewed. All aspects have been discussed in great detail in prior papers [4-8]. 

\begin{figure*}[htp]
    \centering
    {
        \includegraphics[width=8cm]{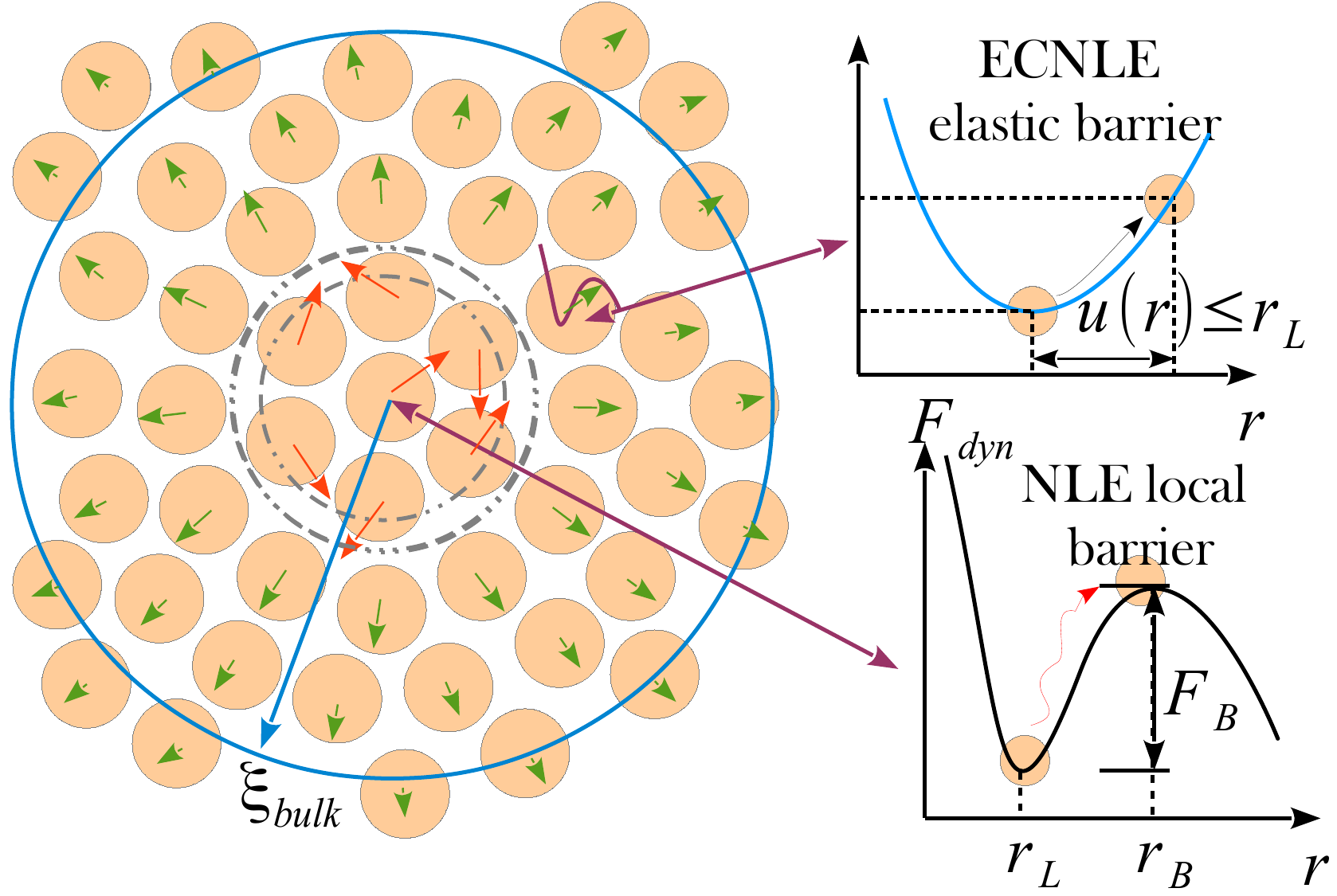}
        \label{fig:first_sub}
    }
    \bigskip
    {
        \includegraphics[width=8cm]{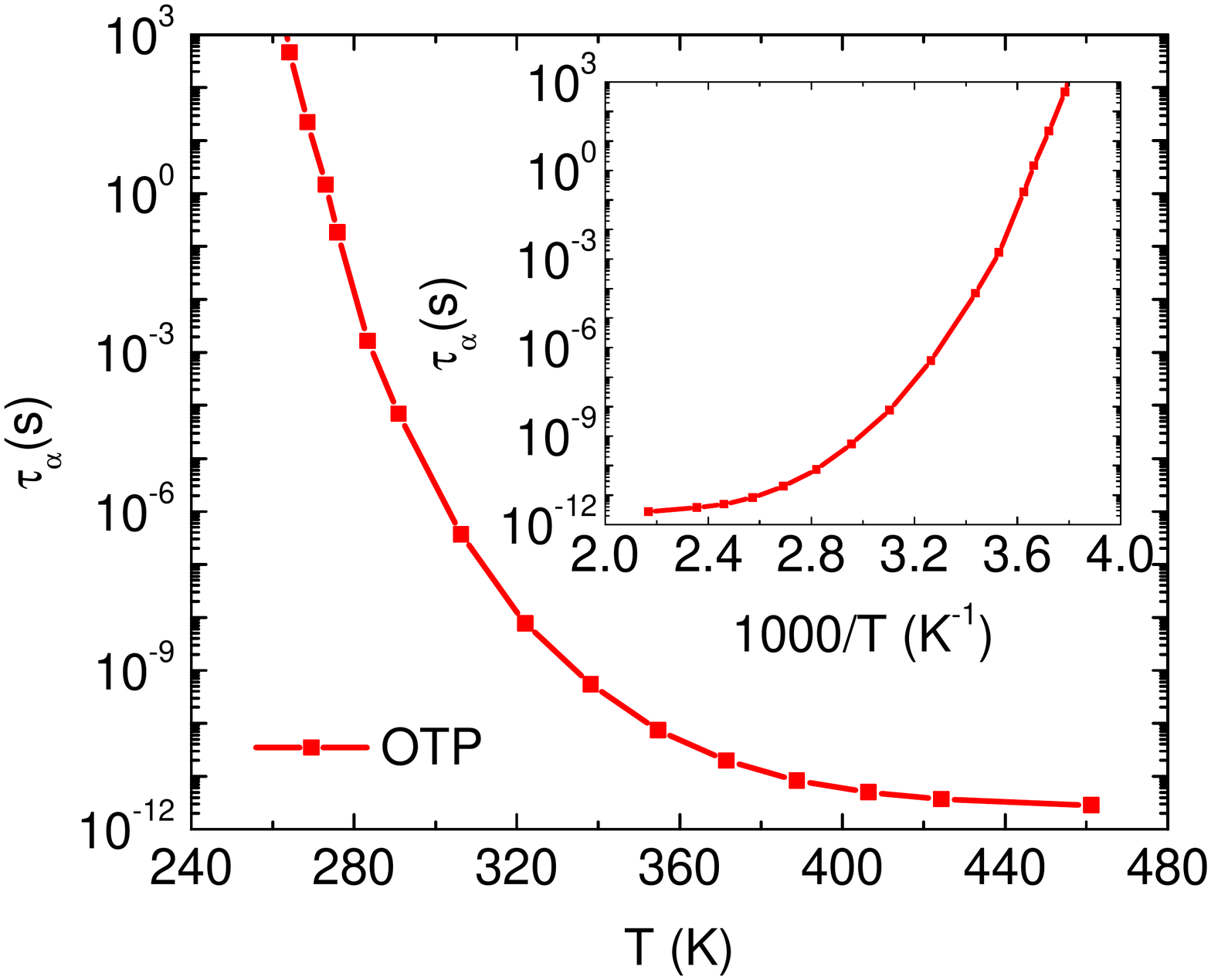}
        \label{fig:second_sub}
    }
    \caption{Left panel. Schematic of the fundamental relaxation event for spheres which involves a local, large amplitude cage-scale hopping motion on the $\sim 3$ particle diameter length scale and a nonlocal, spatially long-range collective elastic motion to accommodate the local rearrangement. Cage scale hopping is described by the dynamic free energy as a function of particle displacement, which sets the amplitude of the long range elastic displacement field outside the cage. Various key length and energy scales are indicated. Right panel shows numerical calculations of the mean alpha time (secs) for OTP liquid as a function of temperature (main frame) and inverse temperature (inset).}
   \label{fig:1}
\end{figure*}

\subsection{Quasi-Universal ECNLE Theory of Spherical Particle Liquids}
ECNLE theory describes the activated relaxation of a tagged particle as a mixed local-nonlocal rare hopping event [6]. Figure 1 shows a cartoon of the key physical elements. The foundational quantity for a tagged spherical particle (diameter, $d$) liquid of packing or volume fraction $\Phi$ is the angularly-averaged instantaneous displacement (denoted, $r$) dependent dynamic free energy, $F_{dyn}(r)=F_{ideal}(r)+F_{cage}(r)$, the derivative of which determines the effective force on a moving particle due to its surroundings. The “ideal” term $\beta F_{ideal}(r)=-3\ln(r/d)$ favors unbounded diffusion or delocalization. The localizing “caging” contribution. The localizing “caging” contribution $F_{cage}(r)$ is constructed from knowledge of the equilibrium pair correlation function $g(r)$ or structure factor $S(k)$. It captures kinetic constraints on the nearest neighbor cage length scale defined from the location of the first minimum of $g(r)$ ($r_{cage}\approx 1.5d$). Large amplitude local hopping drives irreversible rearrangement, but is strongly coupled to (or "facilitated by") a spatially long-range collective elastic adjustment of all particles outside the cage needed to create the extra space required to accommodate a hop.

Key local lengths (see Fig.\ref{fig:1}) are the minimum and maximum of the dynamic free energy ($r_L$ and $r_B$, respectively), and jump distance $\Delta r=r_B-r_L$; key energies are the local cage barrier height,$F_B$, and harmonic curvature at the dynamic free energy minimum, $K_0$. The precise nature of the elastic fluctuation field ($u(r)$ in Fig.1) required to facilitate a cage scale hop is a priori unknown. As a technical approximation, the liquid outside the cage is treated as a continuum linear elastic material (following Dyre [18]) which allows calculation of the displacement field using continuum mechanics supplemented by a microscopic boundary condition [6]. The so-computed radially-symmetric single particle displacement field decays as an inverse square power law of distance [6]:
\begin{eqnarray}
u(r)=\Delta r_{eff}\frac{r_{cage}^2}{r^2}, \qquad r\geq r_{cage}
\label{eq:1}
\end{eqnarray}

The amplitude is set by the microscopically-determined mean cage expansion length, $\Delta r_{eff}$ [6]:
\begin{eqnarray}
\Delta r_{eff}\approx 3\Delta r^2/32r_{cage} \leq r_L
\label{eq:2}
\end{eqnarray}
where $\Delta r \approx 0.2-0.4d$ and grows with density or cooling. The prefactor of 3/32 in Eq.(\ref{eq:2}) follows from assuming each spherical particle in the cage independently hops in a random direction by $\Delta r$.

There are two ways to then compute the elastic barrier. One could invoke literal continuum mechanics, as done by Dyre in his seminal phenomenological approach [18]. However, in ECNLE theory the local cage and long range collective elastic aspects are intimately related. Given the former is described microscopically, for consistency prior work has invoked a particle-level calculation of the elastic barrier we refer to as "molecular Einstein-like". It corresponds to computing the elastic barrier by summing over all harmonic particle displacements outside the cage region which yields: [6] 

\begin{eqnarray}
F_{elastic} &=& \rho\frac{K_0}{2}\int_{r_{cage}}^\infty dr 4\pi r^2 u^2(r)g(r)\nonumber\\
&\approx& 12K_0\Phi\Delta r_{eff}^2\left(\frac{r_{cage}}{d}\right)^3
\label{eq:3}
\end{eqnarray}
where $r$ is relative to the cage center and $K_0=3k_BT/r_L^2$. Note the long range nature of the integrand in eq 3 which decays as $\sim r^{-2}$, and hence the total elastic barrier converges slowly to its full value with the leading correction scaling as $\sim r^{-1}$. 

The sum of the coupled (and in general temperature and density dependent) local and elastic collective barriers determine the mean total barrier for the alpha relaxation process: 
\begin{eqnarray}
F_{total} = F_B + F_{elastic} 
\label{eq:4}
\end{eqnarray}
The elastic barrier increases much more strongly with increasing density or cooling than its cage analog, and dominates the growth of the alpha time as the laboratory glass transition is approached [6].  A generic measure of the average structural relaxation time follows from a Kramers calculation of the mean first passage time for hopping [6]. For barriers in excess of a few kBT one has: [5,6]
\begin{eqnarray}
\frac{\tau_\alpha}{\tau_s} = 1+ \frac{2\pi(k_BT/d^2)}{\sqrt{K_0K_B}}\exp{\frac{F_B+F_{elastic}}{k_BT}}
\label{eq:5}
\end{eqnarray}
where $K_B$ is the absolute magnitude of the barrier curvature in units of $k_BT/d^2$. The alpha time is expressed in units of a "short time/length scale" relaxation process (cage-renormalized Enskog theory) the explicit formula for which is given elsewhere [6,8]. Physically, it is meant to capture the alpha process in the absence of strong caging defined by the parameter regime where no barrier is predicted (e.g., $\Phi < 0.43$ for hard spheres [19]). The latter condition corresponds to being below the naive mode coupling theory "transition" which in ECNLE theory is manifested as a smooth dynamic crossover [6,19,20].

\subsection{Mappings for Molecular and Polymeric Liquids}
The theory is rendered quantitatively predictive for rigid molecular liquids via a mapping [4,5] to an effective hard sphere fluid guided by the requirement that it exactly reproduces the equilibrium dimensionless density fluctuation amplitude (compressibility) of the liquid [21], $S_0(T) = \rho k_BT\kappa_T$. This long wavelength thermodynamic quantity sets the amplitude of nm-scale density fluctuations, and follows from the experimental equation-of-state (EOS). The mapping relation is: [4] 
\begin{eqnarray}
S_0^{HS}&=&\frac{(1-\Phi)^4}{(1+2\Phi)^2}\equiv S_{0,exp}=\rho k_BT\kappa_T\nonumber\\
&\approx &\frac{1}{N_s}\left(-A+\frac{B}{T} \right)^{-2}
\label{eq:6}
\end{eqnarray}
The first equality employs Percus-Yevick (PY) integral equation theory [21] for hard sphere fluids. The final equality is an accurate analytic description of experimental data derived previously [4]. Temperature enters all 3 factors in  $S_0(T)$. This mapping determines a material-specific, temperature-dependent effective hard sphere packing fraction, $\Phi_{eff}(T)$. From Eq.(\ref{eq:6}) one has the explicit expression:
\begin{eqnarray}
\Phi_{eff}(T;A,B,N_s)&=& 1 + \sqrt{S_0^{expt}(T)} \nonumber\\
&-& \sqrt{S_0^{expt}(T) + 3\sqrt{S_0^{expt}(T)}}
\label{eq:7}
\end{eqnarray}

Thus, in practice, 4 known chemically-specific parameters enter in the minimalist mapping [4,5,7,8]: $A$ and $B$ (interaction site level entropic and cohesive energy EOS parameters, respectively), the number of elementary sites that define a rigid molecule, $N_s$ (e.g., $N_s=6$ for benzene), and hard sphere diameter, $d$. Knowledge of $\Phi_{eff}(T)$ allows $g(r)$ and $S(k)$ to be computed, which determines $F_{dyn}(r)$, from which all dynamical results follow. With this mapping, ECNLE theory can make alpha time predictions with no adjustable parameters. The theory has accurately predicted the alpha time over 14 decades for nonpolar organic molecules, and with less quantitative accuracy for hydrogen-bonding molecules (e.g., glycerol).[4,5] 

Figure 1 shows mean alpha relaxation time calculations for orthoterphenyl (OTP) in two temperature representations; for this system no adjustable parameter agreement with experiment has been documented [4,5]. Detailed analytic and numerical analyzes of the theoretical form of the temperature dependence of the alpha time have been performed [4]. Over various restricted temperature or time scale regimes, the theory is consistent with essentially all of the diverse forms in the literature including the entropy crisis VFT [3], dynamic facilitation parabolic model [22], empirical two-barrier form [23,24], and MCT critical power law [25]; see ref 4 for a detailed discussion.

Polymers have additional complexities associated with conformational isomerism and chain connectivity. As a minimalist model the polymer liquid is replaced by a fluid of disconnected Kuhn-sized segments modeled as non-interpenetrating hard spheres composed of a known number of interaction sites, $N_s$, and effective hard core diameter [7]. Polymer-specific errors must be incurred based on such a mapping. To address this, a one-parameter non-universal version of ECNLE theory has been developed based on the hypothesis the amount of cage expansion depends on sub-nm chemical (conformational) details that are coarse-grained over in the effective hard sphere description[8]. Nonuniversality enters via a modified jump distance, $\Delta r \rightarrow \lambda\Delta r$, where the constant $\lambda$ is adjusted to \emph{simultaneously} provide the best theoretical description of $T_g$ and fragility on a polymer-specific basis [8]. From Eqs(\ref{eq:2}) and (\ref{eq:3}), this results in $F_{elastic}\rightarrow \lambda^4 F_{elastic}$. Hence, the relative importance of the local versus collective elastic barrier acquires a polymer-specificity.  Very high (very low) fragility polymers correspond to $\lambda$ values greater (smaller) than the universal model value of unity. Hence, within ECNLE theory increasing $\lambda$ and dynamic fragility corresponds to a more cooperative alpha process as defined by the relative importance of the collective elastic contribution to relaxation [8].

In this article, we present representative calculations for a subset of organic molecules and polymer melts previously studied [4,7,8]. Specifically [8], polystyrene (PS; fragility = $m \sim 110$) and orthoterphenyl (fragility $\sim 82$) where $\lambda_{PS}=\lambda_{OTP}=1$, very high fragility ($m \sim 142$) polycarbonate (PC) where $\lambda_{PC}=\sqrt{2}$, and low fragility ($m\sim 46$) polyisobutylene (PIB) where $\lambda_{PIB}=0.47$.

\section{Temperature Dependence of Short and Long Time Dynamics and Effective Volume Fraction}
\subsection{Apparent Plateau Mean Square Displacement}
The single particle mean square displacement (MSD) at intermediate time scales where particles are approximately "transiently localized" is a quantity of interest in simulation [3,26] and experiment (e.g., quasi-elastic neutron scattering [27]). In a log-log plot, the displacement corresponding to the minimum non-Fickian slope of the MSD serves as an objective and practical measure of a "localization length" [3,28]. Consistent with intuition, it has been numerically shown based on stochastic trajectory solution of NLE theory [29] that this condition corresponds to the mean displacement, $R^*$, where the cage restoring force of the dynamic free energy is a maximum. Analytic analysis of NLE theory yields [30]:
\begin{eqnarray}
R^* \varpropto \sqrt{d.r_L}
\label{eq:8}
\end{eqnarray}

This practical measure of a dynamic localization length is not the same as the literal minimum of the dynamic free energy at $r_L$.

Calculations of $R^*$ as a function of temperature for several systems are shown in Fig.2. Possible universality based on a high temperature crossover temperature, $T_A$, is explored where the latter is defined via when the total barrier is either 1 or 3 $k_BT$. The doubly normalized plot in Fig.2 reveals that over a very wide range of reduced temperatures (corresponding to the alpha time changing by more than 10 decades), a good collapse is found which depends little on chemistry or which criterion is adopted for $T_A$. . Using a prior analytic result [19,30] of NLE theory that $r_L\approx 30d\exp(-12.5\Phi)$ plus Eq.(\ref{eq:8}) one has:
\begin{eqnarray}
\left(\frac{R^*}{R_A^*} \right)^2 \approx \frac{r_L}{r_{L,A}}\approx \exp(-12.5(\Phi(T) - \Phi(T_A)) )
\label{eq:9}
\end{eqnarray}
These results can be potentially tested against simulation and experiment. Note that while the numerical data in Figure 2 can be reasonably described as linear in temperature over the narrow range probed in simulation, the functional form is nonlinear at the lower temperatures of primary experimental interest. This cautions against linear extrapolation of high temperature simulation data.

\begin{figure}[htp]
\includegraphics[width=8cm]{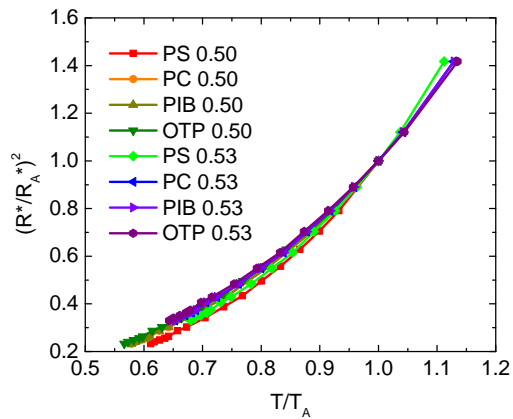}
\caption{\label{fig:2}(Color online) Square of the displacement of maximum cage restoring force normalized by its value at the high temperature reference state as a function of reduced temperature for PS, OTP, PIB, PC and two choices of $T_A$ corresponding to the low barrier states $\Phi_A=0.50$ and $\Phi_A=0.53$. The curves through the points are guides to the eye. }
\end{figure}

\subsection{Dynamic Barriers}
  Fundamental connections of the alpha relaxation time and measures of short time dynamics have been predicted by ECNLE theory in prior studies [4,6]. Recently, Simmons [31,32] suggested based on simulations performed at relatively high temperatures that a roughly exponential, but non-universal, connection exists between the effective barrier deduced from the logarithm of the alpha time and the MSD in the pseudo-plateau regime if both quantities are non-dimensionalized by a high temperature crossover value. In the simulations, the latter is defined as the temperature $T_A$ where there is $\sim 10\%$ deviation from Arrhenius relaxation. 

Motivated by the above, Figure 3 plots our calculations for the temperature-dependent total barrier divided by its value at $T_A$ against the normalized square of $R^*$. For PS, PC, and OTP (not plotted, identical to PS) they are well fit (including all of the deeply supercooled regime) by: 

\begin{eqnarray}
\frac{F_{total}(T)}{F_{total}(T_A)} = -a + b\exp\left[c\left(\frac{R_A^*}{R^*} \right)^2 \right]
\label{eq:10}
\end{eqnarray}
where $a$, $b$, $c$ are positive system-specific constants, and c increases monotonically with fragility. Thus ECNLE theory does predict a specific exponential connection between the barrier and $R^*$ if expressed in a dimensionless form. Note that for the very low fragility PIB ($m\sim 46$), the plot is nearly linear up to a barrier of $\sim 10k_BT$.

\begin{figure}[htp]
\center
\includegraphics[width=8cm]{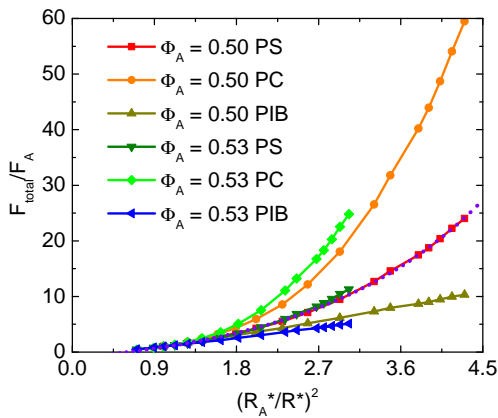}
\caption{\label{fig:3}(Color online) Total dynamic barrier divided by its value at the high temperature reference state versus $(R_A^*/R^*)^2$ for PS, PC, PIB at two values of the high temperature reference state volume fraction. The dotted curve is a fit to the red data points: $3.85+2.77\exp(0.54(R_A^*/R^*)^2)$.}
\end{figure}

Various theoretical models based on different physics generally correspond to different forms of the temperature dependence of the effective barrier [37]. Mauro et al [33] have proposed a phenomenological model they claim can fit experimental data over many decades based on a configurational entropy perspective significantly modified in a manner motivated by constraint theory ideas typically employed for network glass-formers. This model corresponds to an effective barrier in thermal energy units (logarithm of the non-dimensionalized shear viscosity [33]) that grows exponentially with a material-specific energy scale divided by the thermal energy. Hence, a structural relaxation time that is roughly a double exponential of inverse temperature. 

With the above motivation, the main frame of Figure 4a plots ECNLE theory barrier calculations for OTP in a log-linear inverse temperature Angell representation. The individual contributions to the total barrier are not very exponential in inverse temperature. However, surprisingly, the total barrier over a wide range of barrier heights, including the deeply supercooled regime (total barrier ~ 6-32 kBT, corresponding to alpha times ~ 10 ns-100 s) is not far from an apparent Arrhenius form. This seems to us at least partially accidental, given the different physical processes underlying the local and collective elastic barriers in ECNLE theory. The inset of Fig.4a buttresses this view since PS, PC and PIB do not show as good apparent Arrhenius growth of the total barrier as does OTP. 

\begin{figure}[h]
\center
\includegraphics[width=8cm]{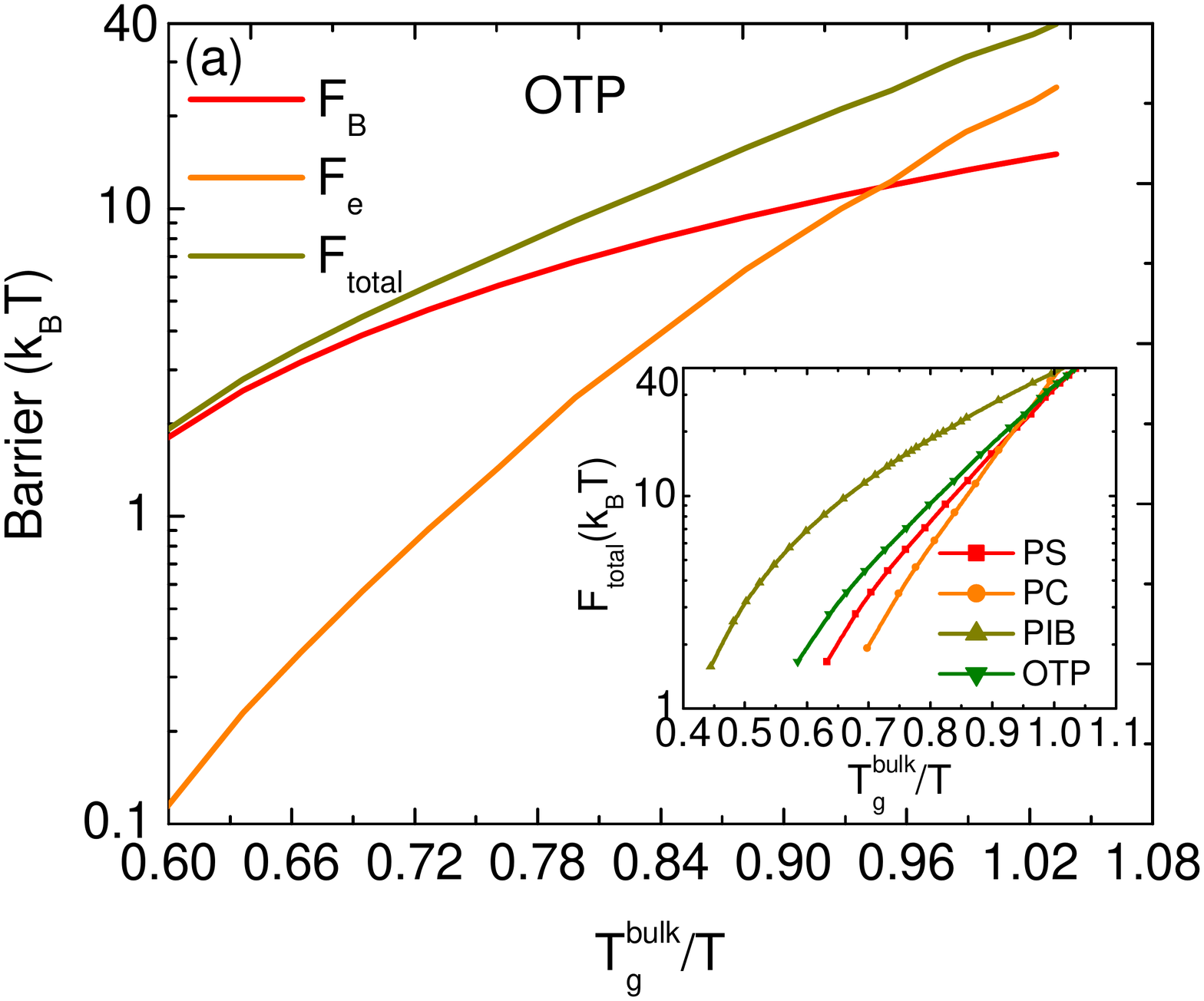}\\
\includegraphics[width=8cm]{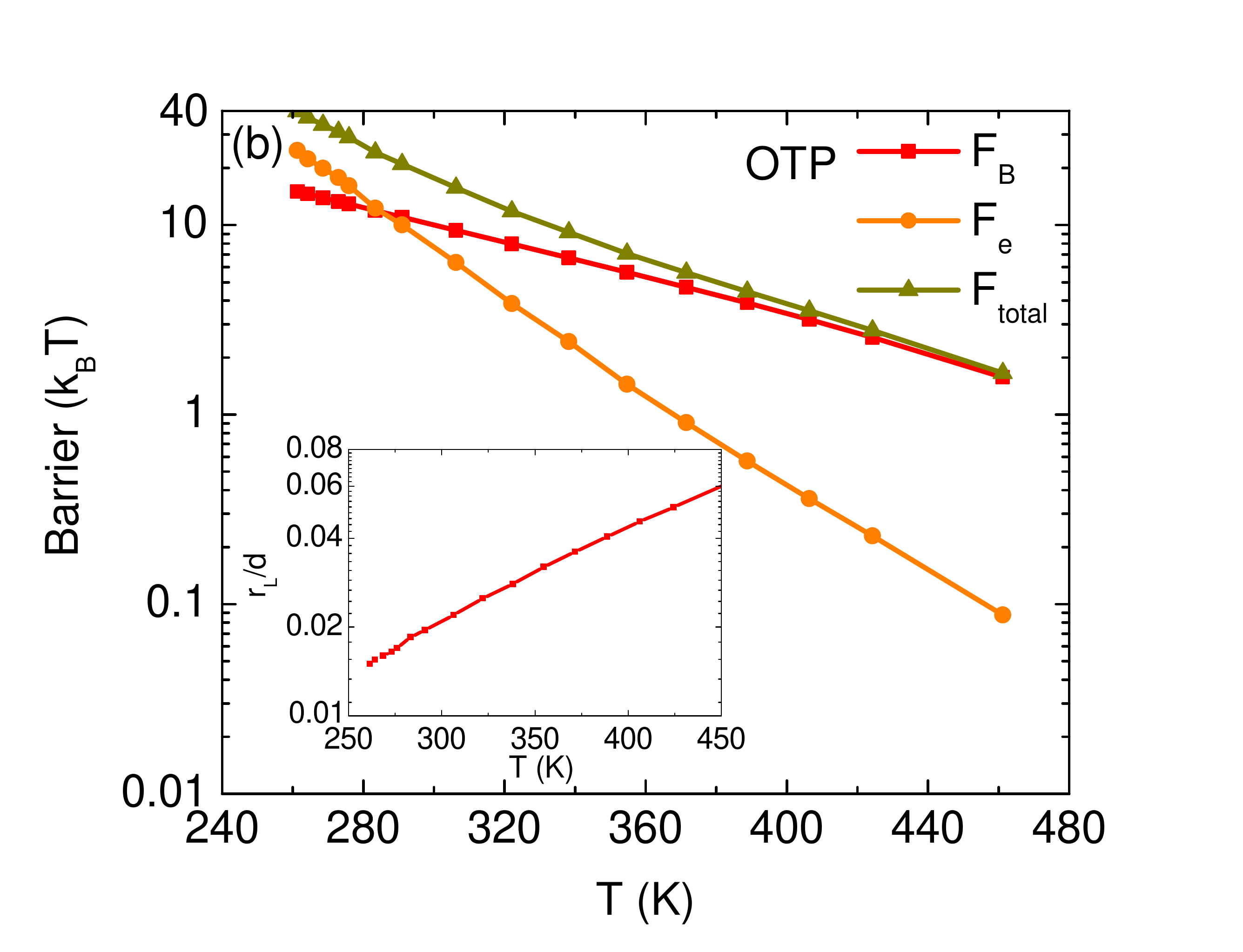}
\caption{\label{fig:4}(Color online) Local, elastic and total barrier as a function of inverse temperature normalized by the bulk $T_g$ for OTP liquid. Inset shows the total barriers for all 4 systems of present interest. (b) OTP results of panel (a)  plotted versus temperature in Kelvin; inset shows the corresponding location of the minimum of the dynamic free energy.}
\end{figure}

Figure 4b plots the same OTP results in the less common linear in temperature format. Curiously, the elastic and total barriers are reasonably exponential in this representation. The inset of Fig.4b shows the corresponding value of $r_L$ for OTP, which is also roughly exponential. This behavior has physical meaning given the connection of $r_L$ with the effective volume fraction in Eq. (\ref{eq:9}), and the near linear growth of the latter with cooling as established in the next sub-section.

\subsection{Mapped Volume Fraction and Dynamic Crossovers}
The key quantity to treat thermal liquids in ECNLE theory is the effective hard sphere temperature-dependent volume fraction of Eq.(7). Figure 5a shows calculations of this quantity for the four systems of present interest in the standard inverse temperature representation. The effective volume fraction grows sub-linearly with inverse temperature, which is perhaps not unexpected given Eqs. (6) and (7). Figure 5b plots the same results versus temperature. Rather surprisingly, the behavior is remarkably simple, following an almost linear growth over a huge temperature range corresponding to a total barrier growth from  $\sim 1-32k_BT$
\begin{eqnarray}
\Phi_{eff}(T)\approx 0.5 + K(T_{ref}-T)
\label{eq:11}
\end{eqnarray}
where $K$ and $T_{ref}$ depend on material. This implies that if the quantities that enter Eq. (\ref{eq:7}) are expanded through linear order in $T$, then the content of the mapping is almost fully captured. We note the slope for OTP in Fig.5b is $\sim 6\times 10^{-4}$ $K^{-1}$, nearly identical to its linear expansion coefficient of $\sim 7\times 10^{-4}$ $K^{-1}$. Precise agreement should not be expected since the mapping is based on the dimensionless compressibility which has 3 temperature dependent quantities. On the other hand, the naive idea that under the isobaric (1 atm) conditions of interest the temperature dependence of $\Phi_{eff}(T)\equiv \rho(T)d_{eff}^3(T)$ is mainly due to thermal expansion (an EOS property) seems reasonable. 

\begin{figure}[htp]
\center
\includegraphics[width=8cm]{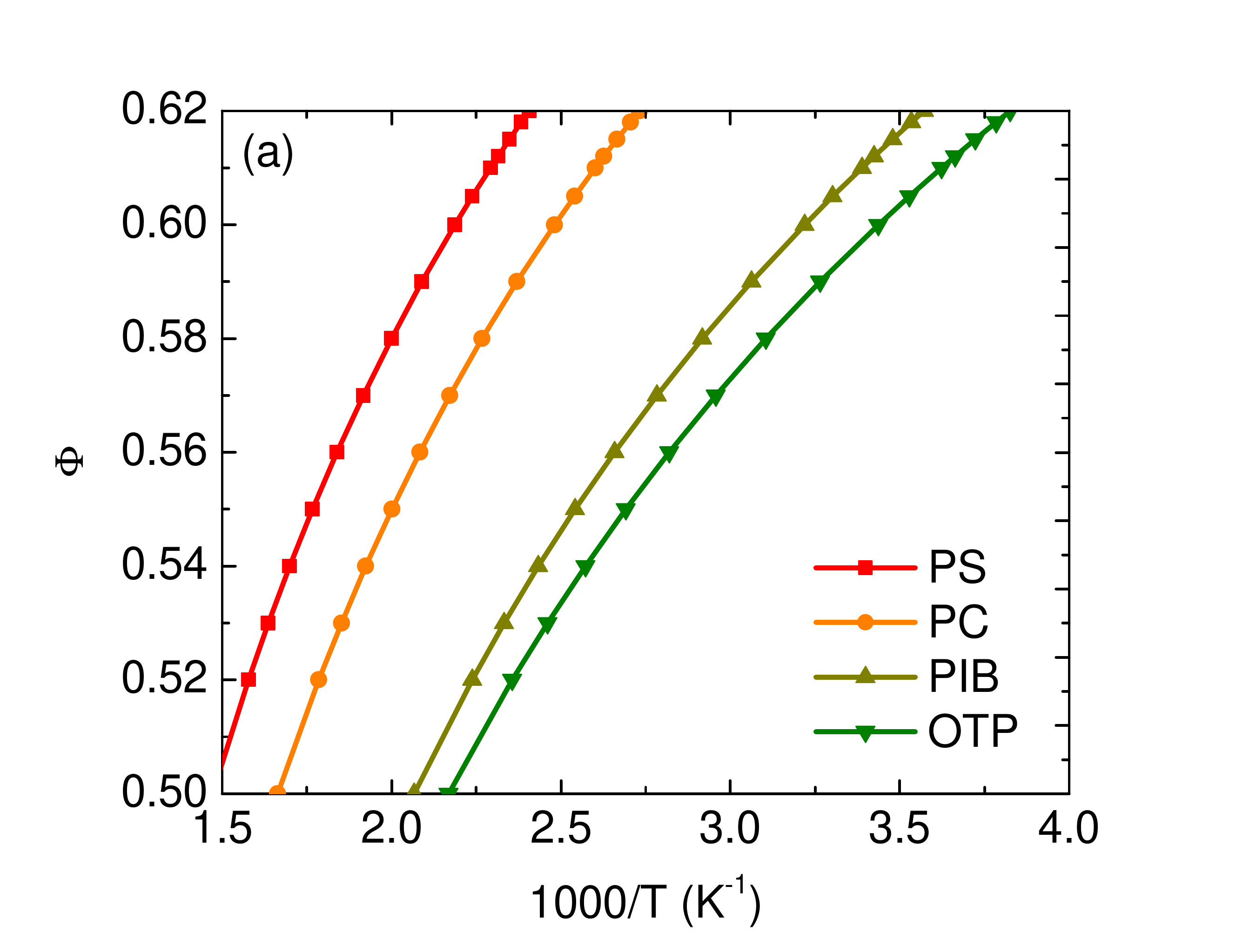}\\
\includegraphics[width=8cm]{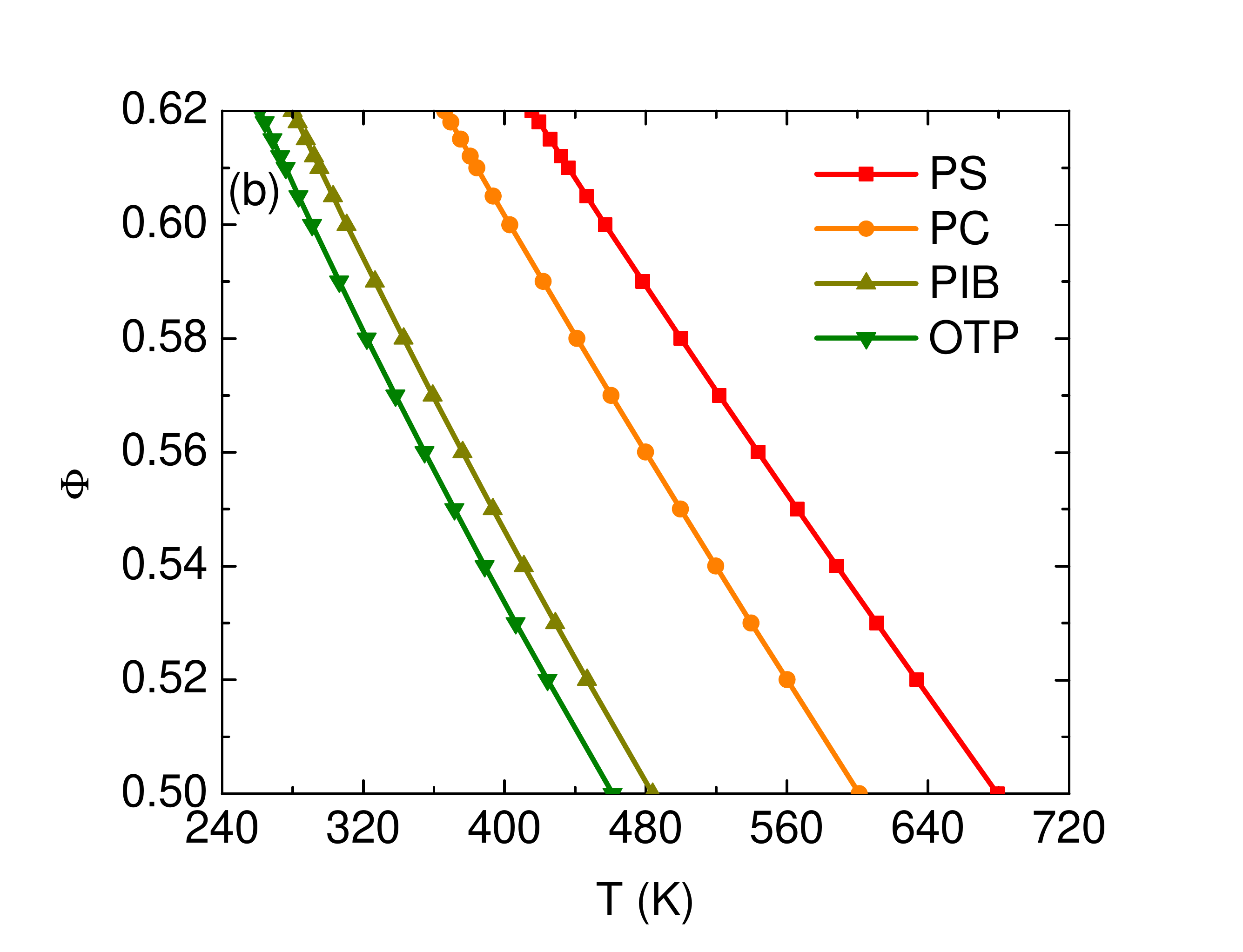}
\caption{\label{fig:5}(Color online) Effective hard sphere volume fraction versus (a) $1000/T$ and (b) $T$ for the 4 systems of interest based on the thermal mapping.}
\end{figure}

The implications of Eq.(\ref{eq:11}) for dynamics are interesting. First note that in the standard representation of Fig.5a the smooth curve could be crudely viewed as consisting of high and low temperature linear branches. With such a construction (not shown), for PS we find the lines intersect at $T^*\sim 540$ $K$, corresponding to $T^*/T_g\sim 1.25$. A similar exercise for OTP and PC yields $T^*/T_g\sim 1.17$ and 1.3, respectively. The absolute value of $T^*/T_g$, and its reduction with increasing fragility, agrees well with trends of experimentally-deduced dynamical crossover temperatures [3,34,35,36]. The latter are based on empirically fitting the ideal MCT critical power law or other functions to alpha time data plotted as a function of inverse temperature. Thus, empirically, one is tempted to associate the smooth thermodynamic $T^*$ crossover as underpinning the dynamical crossover. This perspective in reinforced by examining dynamical properties. A representative example for PS is shown in Fig.6. In either the $T$ or $T^{-1}$ plotting formats, a crossover in the local barrier is found at $\sim 520-540$ $K$, nearly identical to the $T^*$ value found from Fig.5a. A caveat is that although dynamic properties plotted versus $T$ show the crossover, the effective volume fraction of Fig.5b does not.

\begin{figure}[htp]
\center
\includegraphics[width=8cm]{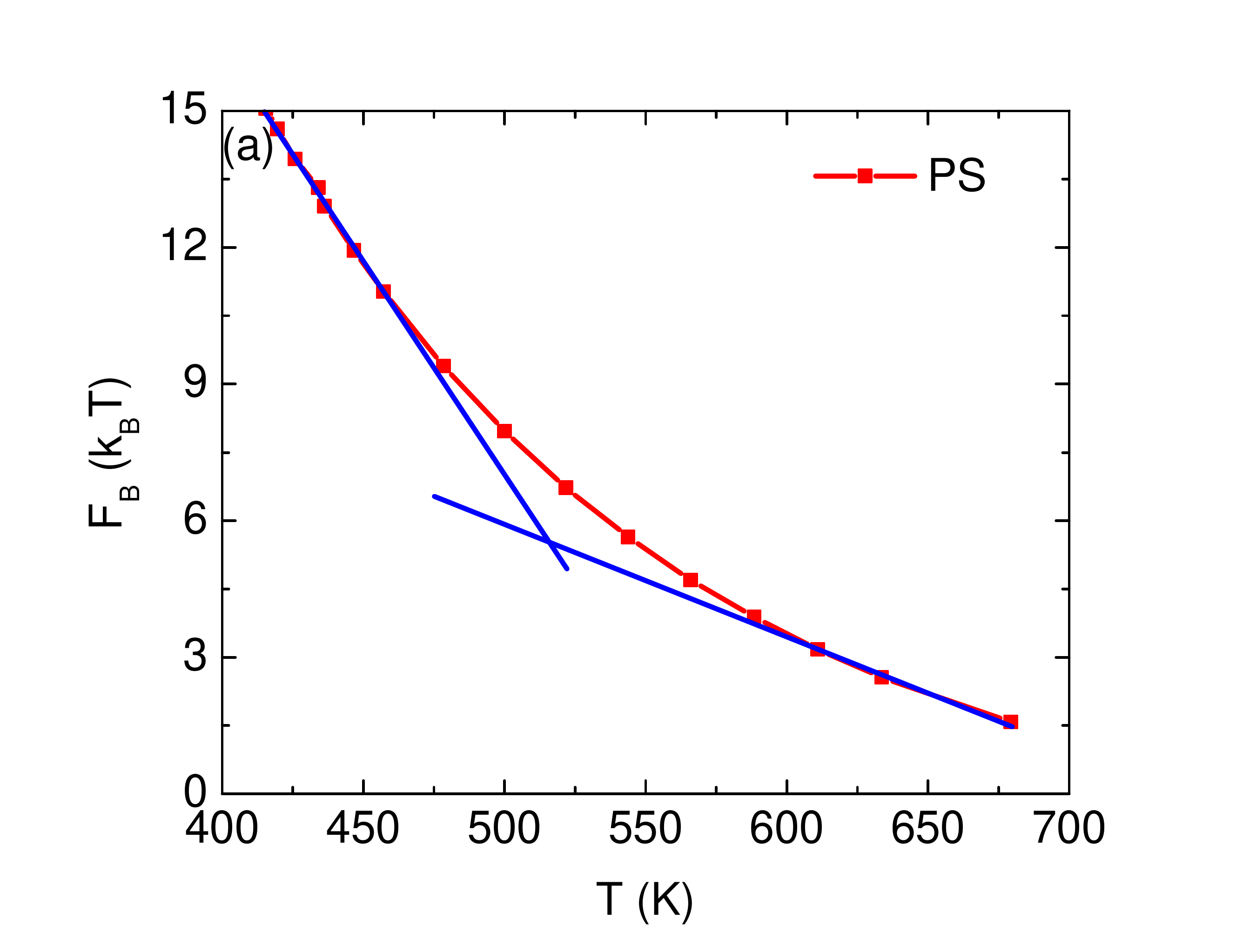}\\
\includegraphics[width=8cm]{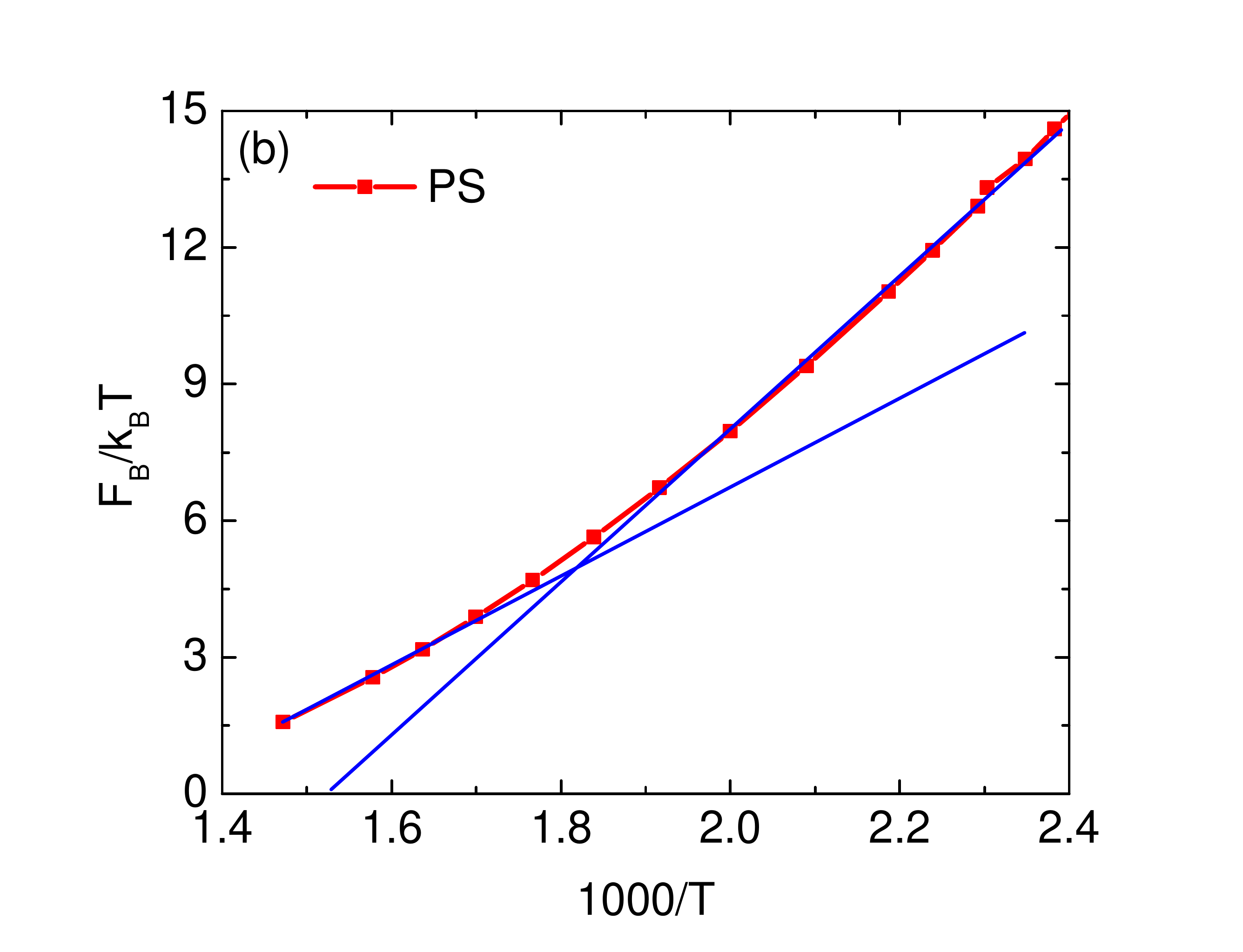}
\caption{\label{fig:6}(Color online) Local cage barrier versus (a) $T$ and (b) $1000/T$ for PS melt. The two blue straight lines show two roughly linear regimes, and their extrapolated intersection occurs at $\sim 515$ $K$.}
\end{figure}

\section{Measures of Cooperativity}
At present, ECNLE theory focuses on average dynamic properties. Explicit space-time dynamic heterogeneity (DH) is not addressed. However, the concept of "cooperativity" is not the same as DH. It can be analyzed in the ECNLE theory framework.

\subsection{Cooperative Displacement}
Real space analyses of simulations [37,38] have attempted to identify the number of particles involved in an relaxation event and, even more objectively, the total particle mean square displacement associated with a re-arrangement, defined here as $MSD^*$. Schall, Spaepen and Weitz [39] experimentally extracted the full displacement field associated with activated re-arrangements in glassy colloidal suspensions. They found a picture akin the ECNLE theory where $\sim 12$ particles in a compact "cage" region of space move by a large amount and are surrounded by a long range collective displacement field. From the observed displacements, the total particle MSD can potentially be measured. 

Based on the coupled local-nonlocal physical picture of alpha relaxation in ECNLE theory, the "number of re-arranging particles" is ill-defined (in contrast to other models such as Adams-Gibbs [40] and RFOT [41] which involve compact clusters). However, we can compute $MSD^*$. The cage consists of a central particle plus $\sim 12$ nearest neighbors. In NLE theory, each particle is envisioned to move a distance $\Delta r$ during the alpha relaxation event. The jump distance increases from $\sim 0.25 - 0.35$ particle diameters upon cooling from the lightly supercooled regime to $T_g$ [6]. Hence, the local component is:
\begin{eqnarray}
MSD^*_{cage}\approx 13\Delta r^2 \approx (0.8-1.6)d^2
\label{eq:12}
\end{eqnarray}
The collective elastic fluctuation contribution corresponds to a total displacement of:
\begin{eqnarray}
MSD^*_{elastic} &=&4\pi\rho\int_d^{\infty}dr r^2\left[\Delta r_{eff}\frac{r_{cage}^2}{r^2} \right]^2 \nonumber\\
&=& 24\Phi\left(\frac{r_{cage}}{d} \right)^4\left(\frac{3\Delta r^2}{32r_{cage}d} \right)^2d^2 \\
&\approx& 0.71\left(\frac{\Delta r}{d} \right)^4\Phi d^2 \approx (0.003-0.011)\Phi d^2 \nonumber
\label{eq:13}
\end{eqnarray}
This is far smaller than the local hopping contribution. Hence, the total linear displacement $\sqrt[]{MSD^*_{total}}$ $\sim 1-2$ particle diameters, grows weakly with cooling, and is dominated by local physics even though the long range elastic effects make a large contribution to the activation barrier. The obtained modest value of $\sqrt[]{MSD^*_{total}}$ does seem reasonable compared to simulation studies [37,38,42].

\subsection{Cooperativity Length Scale}
Since collective elastic effects involve a scale-free displacement field, there is no intrinsic length scale in the usual sense.  However, a cooperativity length can be defined by asking a question recently explored in studies of thin film heterogeneous dynamics [43,44,45,46]. There, one can define a length scale as the distance from the surface where some pre-determined fraction of the bulk alpha relaxation time is recovered. The analog of this idea in the context of bulk ECNLE theory corresponds to adjusting the upper limit in Eq.(3) to define a length-scale-dependent elastic barrier within a spherical region of radius varying from $r_{cage}$ to $\xi_{bulk}$:
\begin{eqnarray}
F_{elastic}(\xi_{bulk}) &=& 4\pi\rho\int_{r_{cage}}^{\xi_{bulk}}dr r^2 g(r)\left[\frac{K_0u^2(r)}{2} \right]\nonumber\\
&=& F_{elastic}^{bulk}\left[1-\frac{r_{cage}}{\xi_{bulk}} \right]
\label{eq:14}
\end{eqnarray}
Note the slow inverse in distance decay to its asymptotic value. From this, a cooperativity length scale is defined as when a fixed percentage ($C$) of the bulk alpha time is recovered:
\begin{eqnarray}
\ln\left(\frac{\tau_\alpha(\xi_{bulk})}{\tau_\alpha^{bulk}} \right) &\equiv& \ln C\approx\frac{F_{elastic}(\xi_{bulk})-F_{elastic}^{bulk}}{k_BT} \nonumber\\
&=& -\frac{r_{cage}}{\xi_{bulk}}\frac{F_{elastic}^{bulk}}{k_BT},
\label{eq:15}
\end{eqnarray}
where $\xi_{bulk}$ is proportional to the bulk elastic barrier. Prior work argued fragility is dominated by collective elasticity which is the origin of "cooperativity" [4,6,8] in ECNLE theory. From Eq.(15) one can write:
\begin{eqnarray}
\xi_{bulk} \approx -\frac{r_{cage}}{\ln C}\frac{F_{elastic}^{bulk}}{k_BT}=-\frac{r_{cage}}{\ln C}\frac{12\Phi K_0\Delta r_{eff}^2}{k_BT}\left(\frac{r_{cage}}{d} \right)^3.
\label{eq:16}
\end{eqnarray}

Figure 7 shows sample calculations of bulk for PS and OTP (they are almost identical) based on the criteria $C=0.5$ and $0.8$. This cooperativity length grows strongly with cooling, and is well described by a cubic polynomial. For the 50 $\%$ criterion, bulk $\xi_{bulk} \sim 30d$ at the laboratory $T_g$. The inset of Figure 7 shows the analogous results for the hard sphere fluid. Concerning the large cooperativity lengths in Figure 7, recall that the emergence of the collective elastic barrier as an important effect begins around a crossover volume fraction of $\sim 0.57-0.58$ [6], and here bulk is relatively small. For example, at $\Phi\sim$ 0.58, the inset of Figure 7 shows that bulk $\xi_{bulk} \sim 6d$  for $C=0.5$. To place this value in context, we note that ECNLE theory predicts for PS parameters that at $\Phi \sim 0.58$ the alpha time $\sim$ 200 nsec. This time scale lies in the practical dynamical crossover regime deduced experimentally for fragile liquids ($\sim$ $10^{-7}$ s) [35,36]. Importantly, it is essentially the longest time scale that has been probed in molecular dynamics (MD) simulation. Hence, since existing MD simulations cannot access the deeply supercooled regime where the collective elastic effects become dominant, the molecular cooperativity lengths they can probe are modest, perhaps no more than $\sim 4-6$ particle diameters.

\begin{figure}[htp]
\center
\includegraphics[width=8cm]{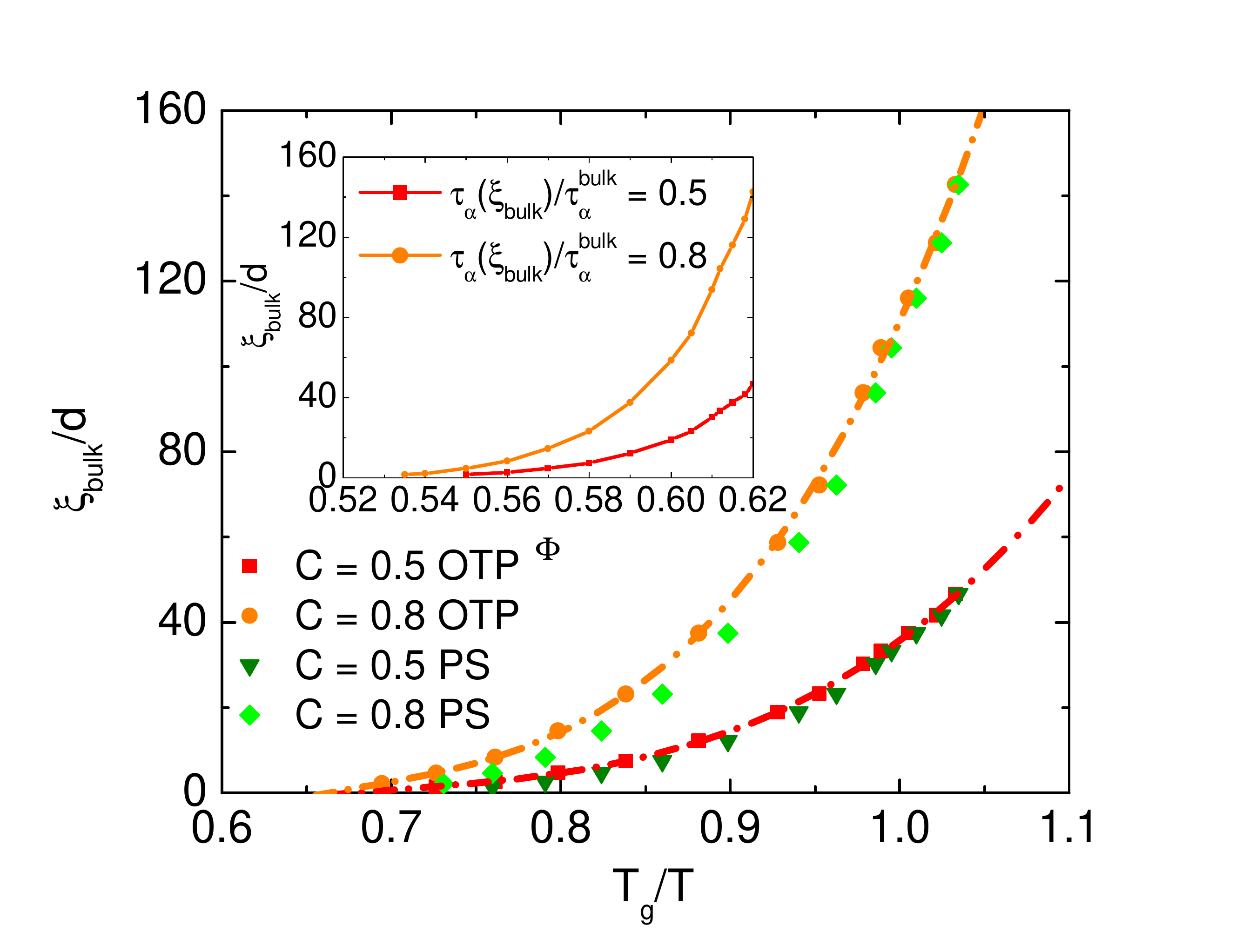}
\caption{\label{fig:7}(Color online) Main frame: Cooperativity length scales defined by the criteria $C= 0.5$ and $0.8$ as a function of inverse normalized temperature by the bulk $T_g$ for OTP and PS. Inset shows the underlying hard sphere fluid results. The dash-dot curves in the main frame are fits to a cubic polynominal.}
\end{figure}

\subsection{Time-Length Scale Connection}
One can ask if a simple connection exists between the cooperativity length and alpha time, or its natural logarithm which defines an effective barrier. This question is of prime interest in diverse glass physics theories [3,26,47]. Each theory typically has a (growing) length scale, with a distinct physical meaning or origin, and often posits a specific power law connection between the effective barrier and this length scale via expressions such as:
\begin{eqnarray}
\ln\left(\frac{\tau_\alpha}{\tau_0}\right)  \varpropto \frac{Barrier}{k_BT} \varpropto \left(\xi/d\right)^\nu \varpropto \frac{\left(\xi/d\right)^\nu}{k_BT} \varpropto \left(\frac{\xi/d}{k_BT}\right)^\nu
\label{eq:17}
\end{eqnarray}
Whether ECNLE theory obeys any of the above three relations is not a priori obvious given the many different microscopic quantities that enter the alpha time calculation and the presence of two barriers with distinct density and temperatures dependences.

We have explored the above question for several thermal liquids and choices of the criterion parameter C. Remarkably, we generically find that all forms in Eq.(17) can represent extremely well our results for the alpha time, typically over 12-15 orders of magnitude in time. Figure 8 shows representative results for the plotting format associated with the final proportionality in Eq.(17). The apparent exponent is $3/4$, and works equally well for two different C values and different chemical species. But the first form in Eq.(17) works just as well (not shown), with an exponent only slightly larger of 0.8. Thus, we robustly find a single activated time-length scale relation holds over essentially the entire temperature regime (lightly to deeply supercooled) with an effective barrier scaling in a weakly sub-linear manner with the cooperativity length. Given Eq.(16), one might think this is not surprising. But recall the alpha time and total barrier involves both local and long range elastic contributions which have very different temperature dependences and relative importances that can vary widely for polymers of diverse fragilities. We note the recent interesting finding of simulations [45,46] of free standing thin films that the bulk alpha time varies exponentially with an effective barrier that grows with roughly one power of a length scale that defines the characteristic width of the mobility gradient near the vapor interface.  

\begin{figure}[htp]
\center
\includegraphics[width=8cm]{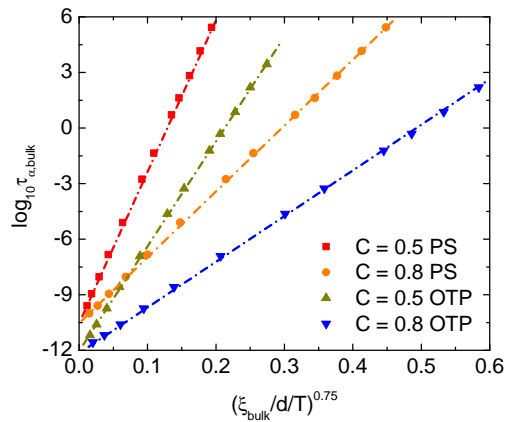}
\caption{\label{fig:8}(Color online) Logarithm of bulk relaxation time as a function of $(\xi_{bulk}/d/T)^{0.75}$ for two systems and two criteria based on the molecular Einstein approach to computing the elastic barrier. Straight dashed lines are fits to the numerical calculations.}
\end{figure}

\section{Alternative Continuum Mechanics Calculation of the Elastic Barrier}
\subsection{Bulk Analysis and Comparison to Einstein Model Analog}
The original motivation for extending the local NLE theory of hopping [6]  to include collective elastic effects was the phenomenological "shoving model" of Dyre [18]. He derived the displacement field in eq 1 albeit with an empirically adjustable amplitude. The elastic energy was computed assuming a literal continuum elastic picture, not the molecular Einstein perspective of ECNLE theory. In our notation, the former corresponds to strain and stress fields in spherical coordinates given by:
\begin{eqnarray}
\varepsilon_{rr}(r)&=&\frac{\partial u(r)}{\partial r}=-\frac{2r_{cage}^2\Delta r_{eff}}{r^3}, \nonumber\\
\varepsilon_{\theta\theta}(r) = \varepsilon_{\varepsilon\varepsilon}(r) &=& \frac{u(r)}{r} = \frac{r_{cage}^2\Delta r_{eff}}{r^3} \\
\sigma_{rr}(r)=-2G\varepsilon_{rr}(r)&,&  \sigma_{\theta\theta}(r)=\sigma_{\varepsilon\varepsilon}(r) = G\varepsilon_{rr}(r) \nonumber
\label{eq:18}
\end{eqnarray}
where $G$ is the high frequency dynamic shear modulus. The strain energy, identified as the elastic barrier, is then:
\begin{eqnarray}
U_e &=&\frac{4\pi}{2}\int_{r_{cage}}^\infty dr r^2\left(\sigma_{rr}(r)\varepsilon_{rr}(r) + 2\sigma_{\theta\theta}(r)\varepsilon_{\theta\theta}(r) \right) \nonumber\\
&=& 8\pi G\Delta r_{eff}^2 r_{cage}
\label{eq:19}
\end{eqnarray}
This basic form is similar to Eq. (\ref{eq:3}) with three differences: (i) numerical prefactor, (ii) the macroscopic shear modulus replaces the single particle spring constant $K_0$, and (iii) the integrand decays not as $r^{-2}$ as does the molecular Einstein model, but much more quickly as $r^{-4}$.

To establish the consequences of the above differences for bulk relaxation, we adopt an accurate analytic formula for G derived in prior NLE theory studies [5,6,30]: 
\begin{eqnarray}
G = \frac{9\Phi k_BT}{5\pi r_L^2d}=\frac{3}{5\pi}\Phi\frac{K_0}{d}
\label{eq:20}
\end{eqnarray}
Substituting Eq.(\ref{eq:20}) into Eq.(\ref{eq:19}) gives 
\begin{eqnarray}
U_e &=& F_{elastic}^{bulk}\frac{2d^2}{5r_{cage}^2}\approx\frac{8}{45}F_{elastic}^{bulk}\nonumber\\ 
&\approx& \frac{F_{elastic}^{bulk}}{6}, \quad \mbox{for} \quad r_{cage}\sim 1.5d
\label{eq:21}
\end{eqnarray}
Hence, almost identical results per the molecular Einstein approach are obtained to within a nearly constant numerical prefactor. For the bulk relaxation time and $T_g$ there are no conceptual differences between using continuum mechanics versus molecular Einstein ideas to compute the elastic barrier.  

\subsection{Cooperativity Length Scale}
\begin{figure}[htp]
\center
\includegraphics[width=8cm]{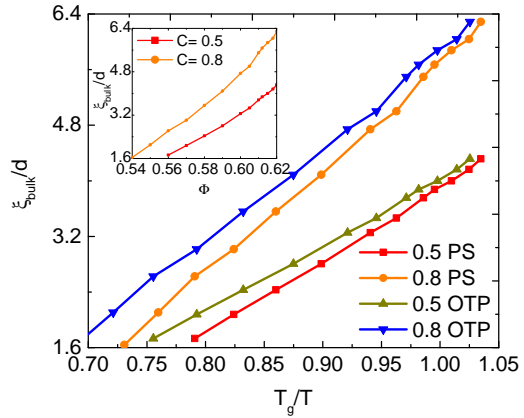}
\caption{\label{fig:9}(Color online) Main frame: cooperativity length scale based on using $C= 0.5$ and $0.8$ as a function of inverse normalized temperature by the bulk $T_g$ for PS liquid based on the continuum mechanics approach to computing the elastic barrier in ECNLE theory. Inset shows the corresponding hard sphere fluid results. }
\end{figure}

Given the elastic energy decays much faster ($\sim r^{-4}$) in the continuum mechanics approach compared to the molecular Einstein analog ($\sim r^{-2}$), there must be significant differences for $\xi_{bulk}$. Figure 9 presents representative calculations analogous to those in Fig.7. One sees a massive reduction in the length scale, and a temperature dependence that is now roughly linear in inverse temperature. The numerical results are easily understood by repeating the analysis in section IVB to obtain:
\begin{eqnarray}
\ln C&\approx& \ln\left(\frac{\tau_\alpha(\xi_{bulk})}{\tau_\alpha^{bulk}} \right) \approx\frac{F_{elastic}(\xi_{bulk})-F_{elastic}^{bulk}}{k_BT} \nonumber\\
&=& -\frac{r_{cage}^3}{\xi_{bulk}^3}\frac{F_{elastic}^{bulk}}{k_BT} \nonumber\\
\ln\left(\tau_\alpha(\xi_{bulk}) \right) &=& \frac{F_B}{k_BT}-\ln C\frac{\xi_{bulk}^3}{r_{cage}^3}.
\label{eq:22}
\end{eqnarray}
Simple algebra yields the relation between the cooperativity lengths based on the two calculations labeled with subscripts $C$ and $E$ for continuum and Einstein, respectively:
\begin{eqnarray}
\frac{\xi_{bulk,C}}{r_{cage}}\approx \left(\frac{\xi_{bulk,E}}{r_{cage}} \right)^{1/3}.
\label{eq:23}
\end{eqnarray}
The cube root relation explains the huge length scale reduction. Given the cubic polynomial fit in Fig. 7, it also explains to zeroth order the nearly inverse temperature dependence in Fig. 9. Note that based on the continuum mechanics calculation, the weakly varying with temperature local barrier now also affects the cooperativity length scale far more than in the molecular Einstein approach. 

\subsection{Time-Length Scale Connection}
We have carried out the same numerical exercise as in section IVC to explore the validity of the three forms of the barrier-alpha time relationships of Eq.(\ref{eq:17}). Results analogous to Fig.8 are shown in Fig.10. Remarkably good straight lines are again obtained, with a much larger apparent exponent now of $\sim 2$. This is roughly three times (as expected given Eq.(\ref{eq:23})) the value of 0.75 found in Fig.8. We also find (not shown) essentially equally good representations of our alpha time calculations by the relations $\ln\left(\tau_\alpha^{bulk} \right) \sim \left(\xi_{bulk}/d \right)^{5/2}$ and $\ln\left(\tau_\alpha^{bulk} \right) \sim \left(\xi_{bulk}/d \right)^{2}/k_BT$.

\begin{figure}[htp]
\center
\includegraphics[width=8cm]{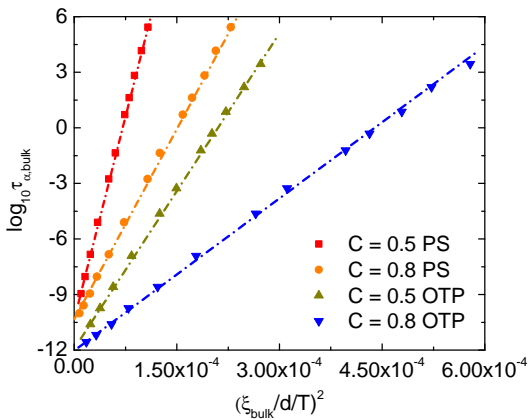}
\caption{\label{fig:10}(Color online) Logarithm of bulk relaxation time as a function of $(\xi_{bulk}/d/T)^2$ for two criteria and two systems based on the continuum mechanics approach to computing the elastic barrier. Straight dashed lines are fits to the numerical data points.}
\end{figure}

We conclude that the existence of a tight connection between the alpha time and a growing cooperativity length scale in ECNLE theory is present regardless of the approach used to compute the elastic barrier. However, the absolute magnitude and temperature dependence of the cooperativity length scale, and the apparent exponent that relates it to the barrier, differ substantially.

\section{Discussion}
We have analyzed new aspects of ECNLE theory to provide deeper insight and address new questions. Calculations have been performed for the hard sphere fluid and thermal molecular and polymeric liquids of diverse fragilities. We find a near universality of the temperature-dependence of the apparent dynamic localization length if one adopts a high crossover temperature as a reference state. In contrast, strong nonuniversalities remain for the total activation barrier. Surprising simplicities emerge for the temperature-dependent effective volume fraction and various dynamical properties if results are plotted against temperature and not its inverse.  

The particle-level total displacement associated with the alpha event is found to be weakly temperature-dependent (grows with cooling) and only $\sim 1-2$ particle diameters. An alternative amplitude-based criterion for determining a cooperativity length scale was also analyzed. It grows strongly with cooling, reaches very large values at the laboratory $T_g$, and is correlated in an exponential manner with the alpha time over an enormous number of decades in relaxation time with a barrier-length scale apparent scaling exponent modestly smaller than unity. An alternative calculation of the elastic barrier based on continuum mechanics results in little change of the predictions of ECNLE theory for bulk average properties, but leads to a much smaller and more weakly growing with cooling cooperativity length scale due to the stronger spatial decay of the elastic field with distance.
    
The issue of the molecular Einstein versus literal continuum mechanics approach to computing  the collective elastic barrier might be more incisively probed by performing new simulations and/or confocal imaging experiments in colloidal materials. This question is especially germane to how solid or vapor boundaries can "cut off" or modify the elastic barrier in thin films [9-12]. Work in this latter direction is underway and will be reported in a future article.  

\begin{acknowledgments}
This work was performed at the University of Illinois and supported by DOE-BES under Grant No. DE-FG02-07ER46471 administered through the Frederick Seitz Materials Research Laboratory. We thank Professor David Simmons for many stimulating and informative discussions.
\end{acknowledgments}

\end{document}